\documentclass[prl,twocolumn,showpacs,preprintnumbers,amsmath,amssymb,superscriptaddress]{revtex4}
\usepackage{graphicx}% Include figure files
\usepackage{dcolumn}% Align table columns on decimal point
\usepackage{bm}% bold math

\begin{document}

\newcommand*{\cm}{cm$^{-1}$\,}
\newcommand*{\Tc}{T$_c$\,}

%\reprint{APS/123-QED}

\title{Three dimensionality of band structure and a large residual
quasiparticle population in Ba$_{0.67}$K$_{0.33}$Fe$_2$As$_2$ as revealed
by the c-axis polarized optical measurement}% Force line breaks with \\

\author{B. Cheng}
\author{Z. G. Chen}

\affiliation{Beijing National Laboratory for Condensed Matter
Physics, Institute of Physics, Chinese Academy of Sciences,
Beijing 100190, China}

\author{C. L. Zhang}
\affiliation{Department of Physics and Astronomy, The University
of Tennessee, Knoxville, Tennessee 37996-1200, USA}

\author{R. H. Ruan}
\author{T. Dong}
\author{B. F. Hu}
\author{W. T. Guo}
\author{S. S. Miao}
\author{P. Zheng}
\author{J. L. Luo}
\author{G. Xu}

\affiliation{Beijing National Laboratory for Condensed Matter
Physics, Institute of Physics, Chinese Academy of Sciences,
Beijing 100190, China}

\author{Pengcheng Dai}

\affiliation{Department of Physics and Astronomy, The University
of Tennessee, Knoxville, Tennessee 37996-1200, USA}

\affiliation{Beijing National Laboratory
for Condensed Matter Physics, Institute of Physics, Chinese
Academy of Sciences, Beijing 100190, China}

\affiliation{Neutron Scattering Science Division, Oak Ridge
National Laboratory, Oak Ridge, Tennessee 37831, USA}

\author{N. L. Wang}

\affiliation{Beijing National Laboratory for Condensed Matter
Physics, Institute of Physics, Chinese Academy of Sciences,
Beijing 100190, China}

\begin{abstract}
We report on a c-axis polarized optical measurement on a
Ba$_{0.67}$K$_{0.33}$Fe$_2$As$_2$ single crystal. We find that the
c-axis optical response is significantly different from that of
high-T$_c$ cuprates. The experiments reveal an anisotropic
three-dimensional optical response with the absence of the
Josephson plasma edge in R($\omega$) in the superconducting state.
Furthermore, different from the ab-plane optical response, a large
residual quasiparticle population down to $T\sim\frac{1}{5}T_c$
was observed in the c-axis polarized reflectance measurement. We
elaborate that there exist nodes for the superconducting gap in
regions of the 3D Fermi surface that contribute dominantly to the
c-axis optical conductivity.

\end{abstract}


% PACS, the Physics and Astronomy
% Classification Scheme.

%\keywords{Suggested keywords}%Use showkeys class option if keyword
                              %display desired
\maketitle

For quasi-two dimensional superconducting materials, striking
difference could exist between the in-plane and out-of-plane (the
c-axis) optical response. In high-T$_c$ cuprates, for example, the
conducting CuO$_2$-planes are usually separated by the insulating
blocking layers. In the normal state, those CuO$_2$-planes appear
to be almost decoupled, leading to the non-metallic charge
transport and dynamics along the c-axis . However, as soon as the
systems enter into the superconducting state, the CuO$_2$-planes
are coupled via the Josephson tunnelling effect, then a sharp
reflectance edge arising from the condensed superconducting
carriers immediately shows up in the c-axis
polarization.\cite{Tamasaku,Basov,Uchida,Shibata,Motohashi} The
plasma edge is referred to as the Josephson plasma edge, which is
linked with the c-axis penetration depth. It offers a direct
measure to the strength of the interlayer coupling. This behavior
is also connected with the mechanism of high-temperature
superconductivity, since the absence of this edge above T$_c$ is
considered as indicative of the confinement mechanism.

Fe-pnictide superconducting materials also crystalize in the
layered structure with FeAs layers separated by alkaline earth
metal ions (Ba$^{2+}$, Sr$^{2+}$) or other Re-O (Re=La, or
rare-earth elements with trivalent Re$^{3+}$ and negative divalent
O$^{2-}$) layers which should be insulator-like. It is important
to see whether or not the Fe-pnictides share similar anisotropic
charge dynamical properties with cuprates. Although there are a
number of optical studies on the ab-plane properties of Fe-based
superconducting materials, little is known about the properties
along the perpendicular direction due to the lack of thick enough
single crystal samples. The c-axis polarized optical data are
available only for parent compounds where the spin-density-wave
gap structure was found to be substantially different from that of
\textbf{E}$\parallel$ab-plane.\cite{ZGChen} In this work we
present the c-axis optical spectroscopy measurements on
superconducting (Ba,K)Fe$_2$As$_2$ crystals. We show that, in
contrast to the high-T$_c$ cuprates, the c-axis Josephson plasmon
is completely absent in the superconducting state. Furthermore,
different from the ab-plane optical response where the full
superconducting energy gaps could be seen clearly \cite{Li}, the
c-axis polarized measurement reveals a very small difference
between T$>$T$_c$ and T$<$T$_c$. The low frequency optical
conductivity below T$_c$ still exhibits a Drude response. The
experiment reveals a large residual quasiparticle population down
to the lowest measurement temperature $T\sim\frac{1}{5}T_c$. Since
the c-axis polarized measurement mainly probes the
three-dimensional (3D) Fermi surface (FS) in the band structure,
the data strongly suggest the presence of nodes for the
superconducting gap in regions of the 3D FS that contribute
dominantly to the c-axis optical conductivity.

Thick Ba$_{0.67}$K$_{0.33}$Fe$_2$As$_2$ single crystals were grown
from the FeAs self-flux in Al$_2$O$_3$ crucibles sealed in Ta
tubes filled with Argon. The same batch of crystals were also used
for neutron scattering experiment, which shows an absence of the
static antiferromagnetic order co-existing with superconductivity
down to 2 K\cite{CLZhang}. The shinny cleaved ab-plane could be
easily obtained after cutting the Ta tube and breaking the
crucible. The layered stacking could actually be seen along the
edge for thick single crystals, making it rather easy to identify
the c-axis. Then the crystals were cut in a direction
perpendicular to the cleaved ab-plane. The cutting surfaces with
dimensions up to 5mm$\times$2mm were finely polished for the
c-axis polarized measurement. The optical reflectance measurements
with \textbf{E}$\parallel$c-axis were performed on a Bruker IFS
80v and 113v spectrometers in the frequency range from 30 to 25000
cm$^{-1}$. An \textit{in situ} gold and aluminium overcoating
technique was used to get the reflectivity R($\omega$). The real
part of conductivity $\sigma_1(\omega)$ is obtained by the
Kramers-Kronig transformation of R($\omega$).

Figure 1 shows the temperature dependence of the in-plane dc
resistivity measured by the four contact technique for a
Ba$_{0.67}$K$_{0.33}$Fe$_2$As$_2$ crystal. The measurement
indicates a superconducting transition temperature T$_c$=38 K. The
resistivity data are very similar to that reported for a
Ba$_{0.6}$K$_{0.4}$Fe$_2$As$_2$ crystal where the in-plane optical
data were presented\cite{Li}.

\begin{figure}
\includegraphics[width=6.5cm,clip]{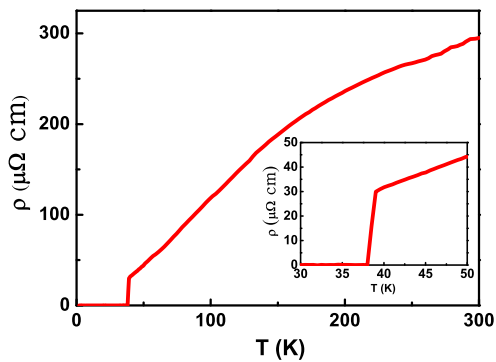}
%\centerline{\includegraphics[width=3.2in]{fig1.eps}}%
\caption{Temperature dependent resistivity $\rho(T)$ for the
Ba$_{0.67}$K$_{0.33}$Fe$_2$As$_2$ single crystal. The
superconducting transition temperature is 38 K.}
\end{figure}

Figure 2 shows the c-axis R($\omega$) and $\sigma_1(\omega)$ at
room T up to 15000 \cm. For a comparison, we also plot the optical
spectra with \textbf{E}$\parallel$ab-plane for the same crystal.
The in-plane optical data are rather close to that of our earlier
measurement on a Ba$_{0.6}$K$_{0.4}$Fe$_2$As$_2$ crystal\cite{Li}.
Similar to the undoped compound, we find that the overall
R($\omega$) along the c-axis is quite similar to that in the
ab-plane except for relatively lower values. This is dramatically
different from the optical spectra of high-T$_c$
cuprates\cite{Uchida} and other layered compounds, for example,
layered ruthenates\cite{Katsufuji}, where the c-axis R($\omega$)
shows much lower values and quite different frequency-dependent
behavior from the ab-plane. This observation suggests that the
band structure of Fe-pnictide should be quite three-dimensional
(3D), in contrast to the expectation based on its layered crystal
structure. An anisotropy ratio of optical conductivities at the
low frequency limit is less than 4, which is close to the value of
the undoped compound \cite{ZGChen}, suggesting that the anisotropy
does not show significant change upon K-doping.

We are mainly interested in the evolution of optical spectra at
low frequencies across the superconducting transition. Figure 3
(a)-(d) show the temperature dependence of the low-$\omega$
reflectance and conductivity spectra. The reflectance R($\omega$)
spectra below 1500 \cm are displayed in Fig. 3(a). The spectra
indicate a typical metallic temperature dependence: the
low-$\omega$ R($\omega$) increases with decreasing temperature.
Most noticeably, no sharp plasma edge develops below T$_c$. This
is in sharp contrast to the high-T$_c$ cuprates where steep
reflectance edge is seen in the superconducting state which was
ascribed to the Josephson coupling of superconducting CuO$_2$
planes\cite{Tamasaku,Basov,Uchida,Shibata,Motohashi}. The result
unambiguously illustrates that the 122-type Fe-pnictides are
significantly different from the cuprates: the condensed
superconducting carriers are not coupled though the Josephson
tunnelling effect. Instead, the metallic optical response
indicates that the compound behaves in a way similar to an
anisotropic 3D system where dispersive bands along the c-axis
would exist.

\begin{figure}[t]
\begin{center}
\includegraphics[width=3.2in]{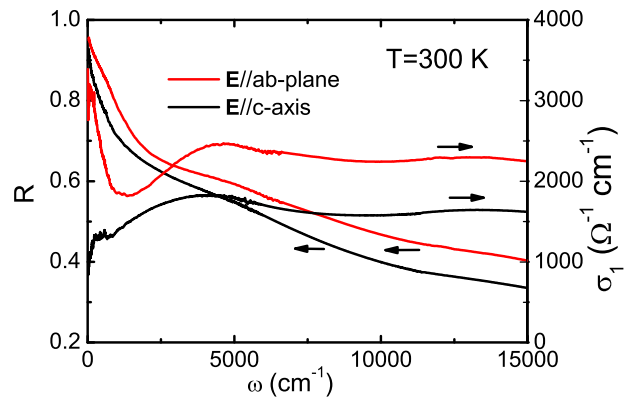}%
%\vspace*{-0.20cm}%
\caption{The c-axis R($\omega$) and $\sigma_1(\omega)$ at 300 K
for the Ba$_{0.67}$K$_{0.33}$Fe$_2$As$_2$ over a broad frequency
up to 15000 \cm. The ab-plane spectra were also included for a
purpose of comparison.}
\end{center}
\end{figure}

Figure 3 (c) shows R($\omega$) at 45K and 8 K in the expanded
region. The R($\omega$) spectrum at 8 K deviates upward from that
at 45 K below roughly 250 \cm, which could be attributed to the
superconducting condensate. However, the difference between the
two spectra is rather small. This is significantly different from
the in-plane optical response. For a comparison, we also plot the
in-plane optical data at the two different temperatures measured
on the same sample (Fig. 3 (e) and (f)). The spectrum at 8 K shows
a clear upward deviation from the spectrum at 45 K below about 300
\cm, and approaches to unity at about 100 \cm, a behavior being
already analyzed in detail for a Ba$_{0.6}$K$_{0.4}$Fe$_2$As$_2$
crystal.\cite{Li} The upward deviation is a characteristic feature
of energy gap opening in the superconducting state. The feature
has also been observed in the in-plane spectra of a number of
other Fe-based materials.\cite{Heumen,Kim,DWu,Homes,Lobo,JTu} The
very subtle difference of the spectra between 45 K and 8 K for
\textbf{E}$\parallel$c indicates the presence of a significant
amount of quasiparticles which have dominant contribution to the
c-axis conductivity.

Figure 3 (b) and (d) show the low-frequency temperature dependence
of the conductivity $\sigma_1(\omega)$ spectra of the sample.
Although the sample shows a metallic temperature dependence of the
low frequency conductivity, the frequency dependence at high
temperature (T$>$45 K) is not simply Drude-like. Instead, a broad
peak at finite frequency exists in the $\sigma_1(\omega)$ spectra.
The peak, which locates near 300 \cm at 300 K, shifts to lower
frequency and gradually disappears with decreasing the
temperature. A Drude component centered at zero frequency is seen
at low temperature. It is worth noting that the presence of the
conductivity peak at finite frequency does not necessarily mean
the absence of the dispersive band. Nevertheless, it is often seen
in poor metallic materials at high temperature where the
quasi-particle peaks along the dispersive band are rather
broad.\cite{NLWang,Valla} Similar to the reflectance spectra, we
can see from Fig. 3 (d) that there is only very small difference
for the conductivity spectra at 8 and 45 K for
\textbf{E}$\parallel$c, while dramatic difference is seen for
\textbf{E}$\parallel$ab-plane (Fig. 3 (f)) which is attributed to
the opening of a superconducting energy gap. The data show that an
s-wave superconducting gap structure could be clearly observed
only from the in-plane optical measurement. The low frequency
optical conductivity still exhibits a Drude response for
\textbf{E}$\parallel$c-axis. The measurement reveals a large
residual quasiparticle population down to the lowest measurement
temperature at 8 K ($T\sim\frac{1}{5}T_c$).

\begin{figure}
\includegraphics[clip,width=1.65in]{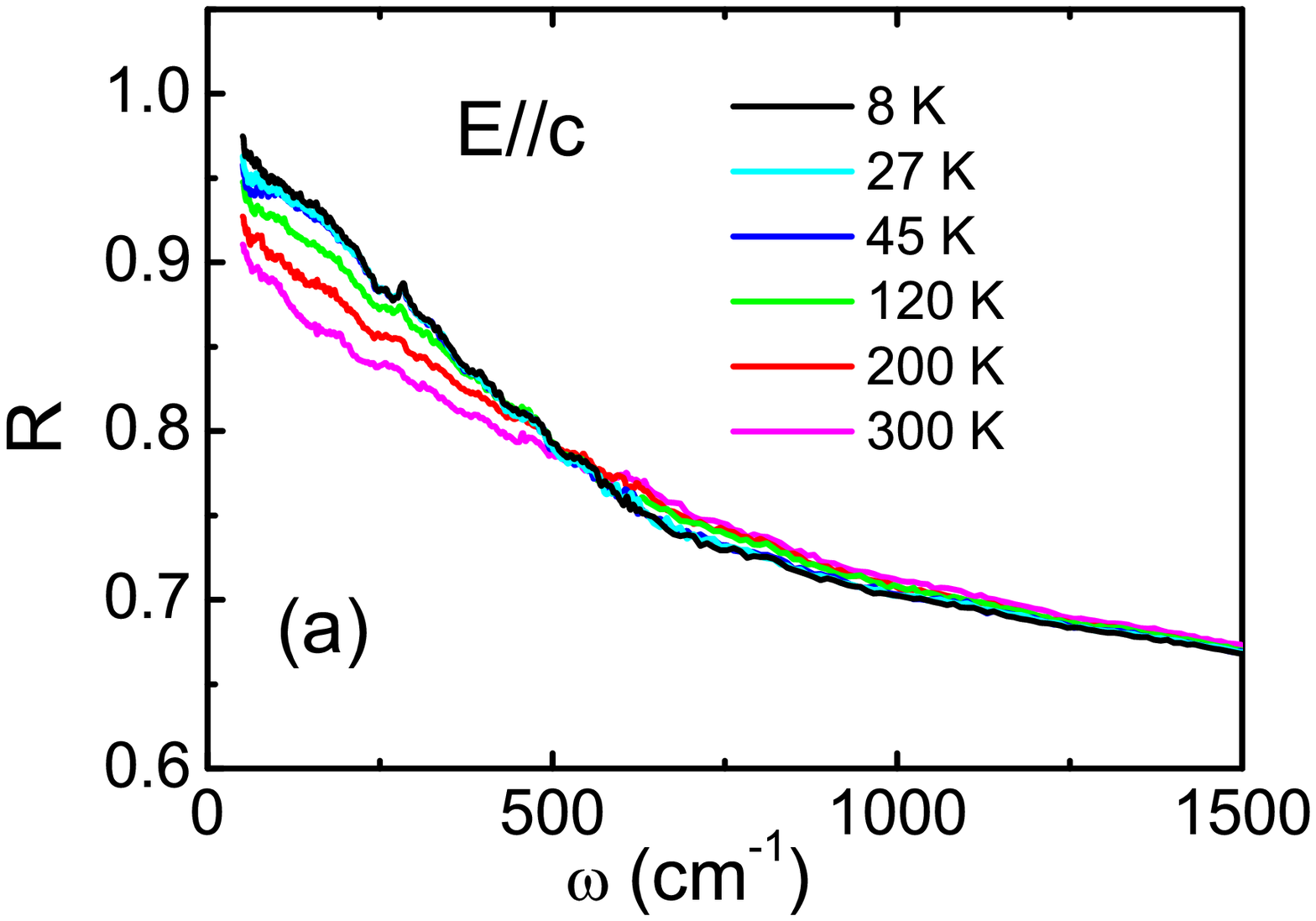}
\includegraphics[clip,width=1.65in]{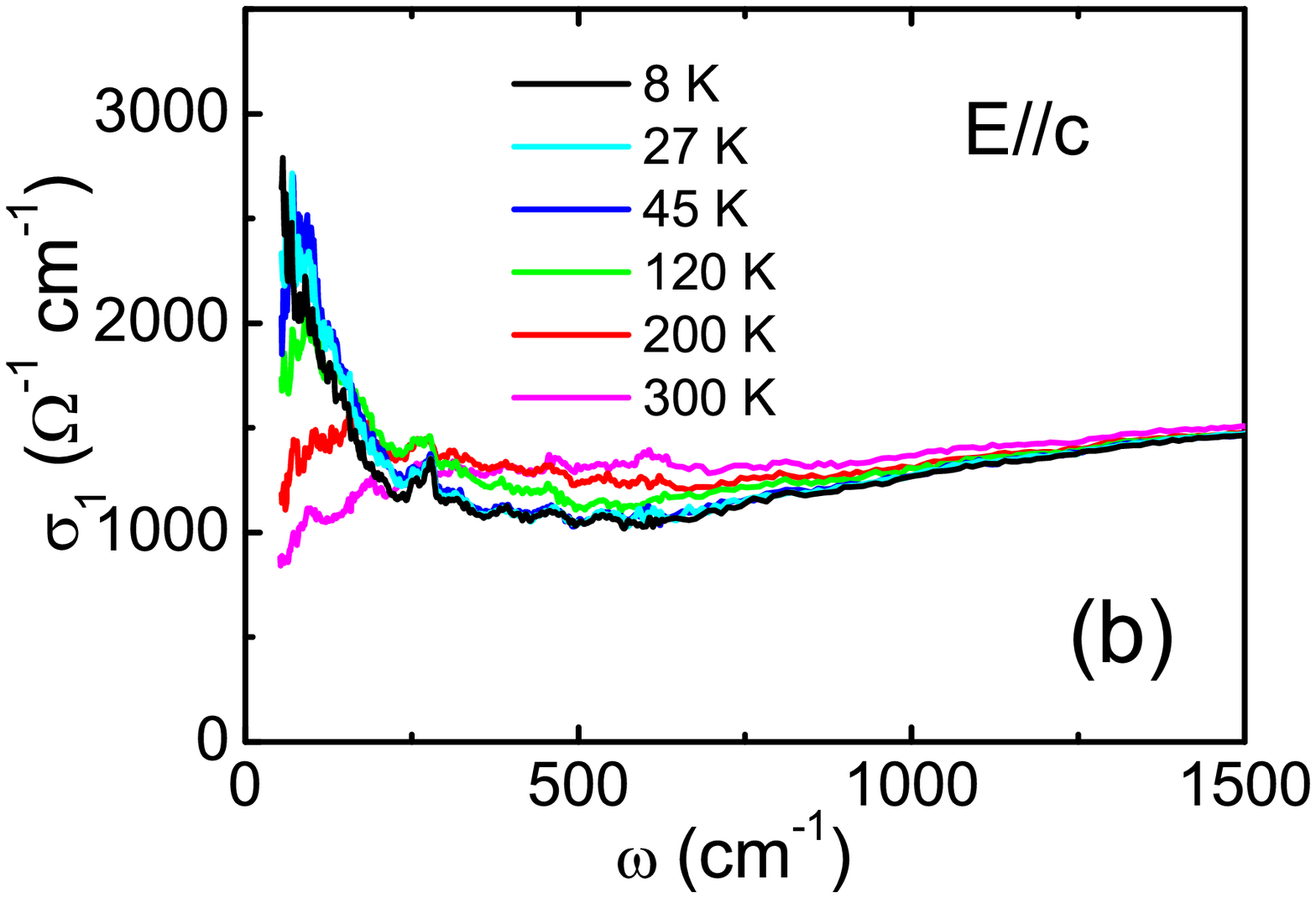}
\includegraphics[clip,width=1.65in]{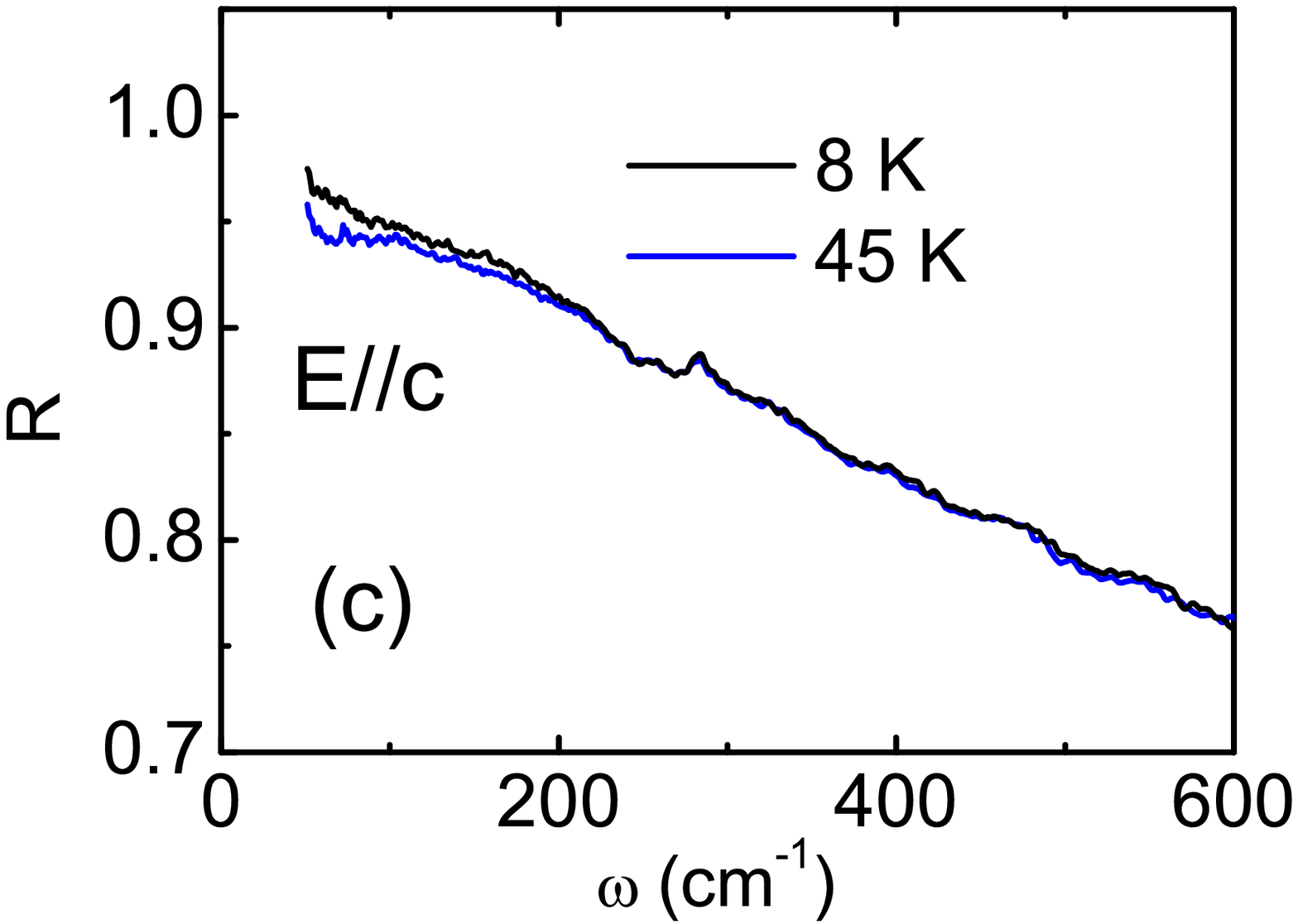}
\includegraphics[clip,width=1.65in]{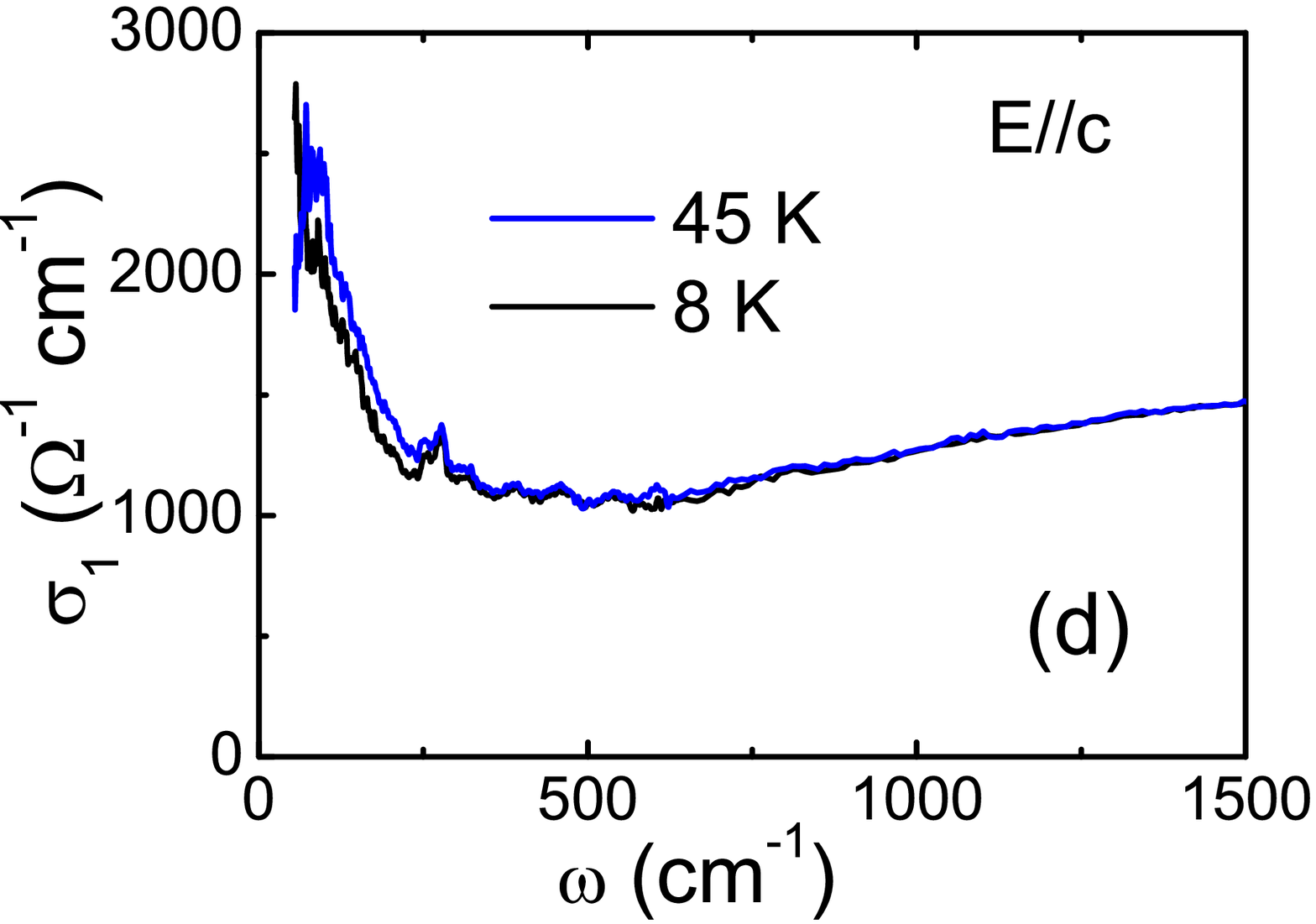}
\includegraphics[clip,width=1.65in]{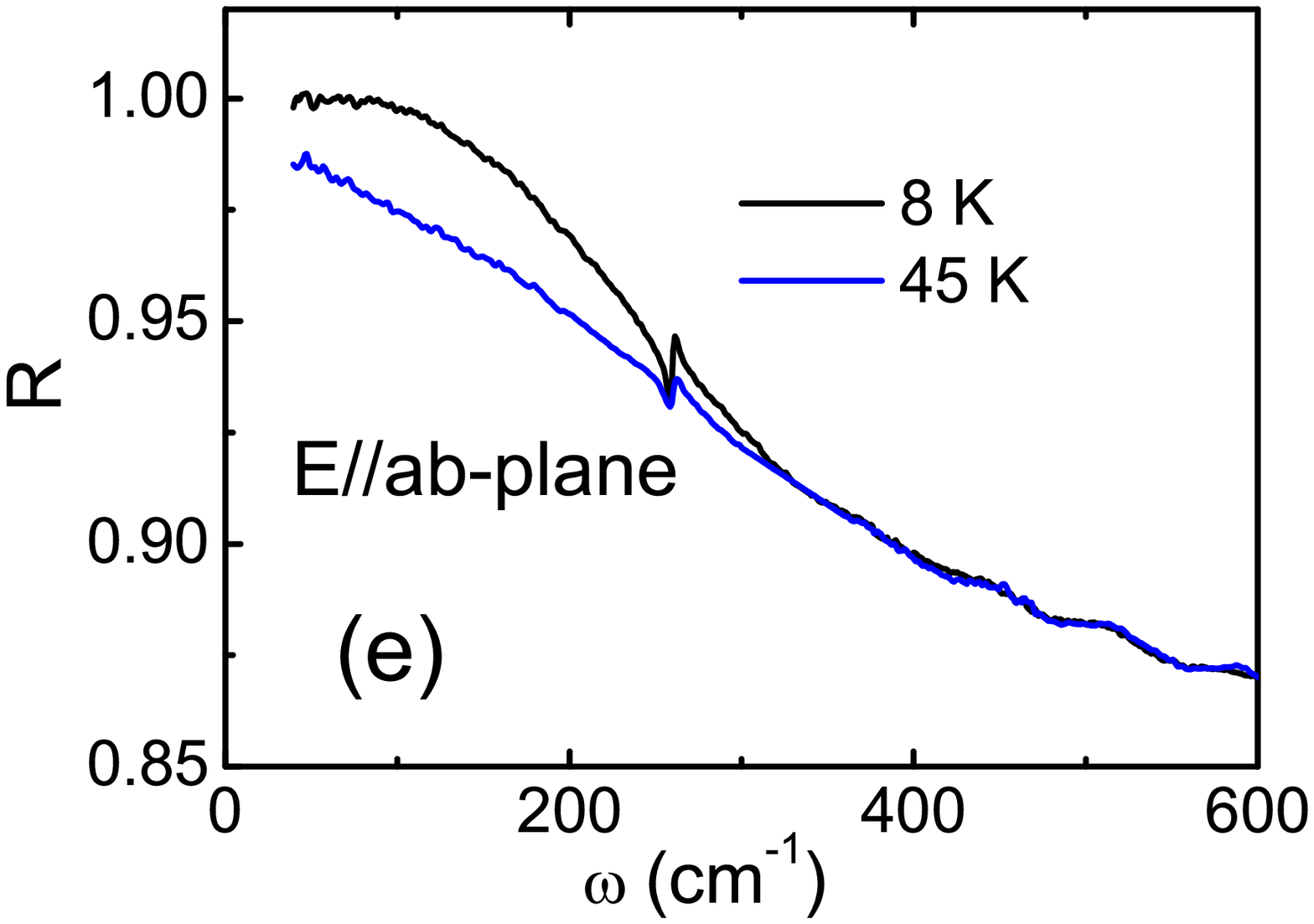}
\includegraphics[clip,width=1.65in]{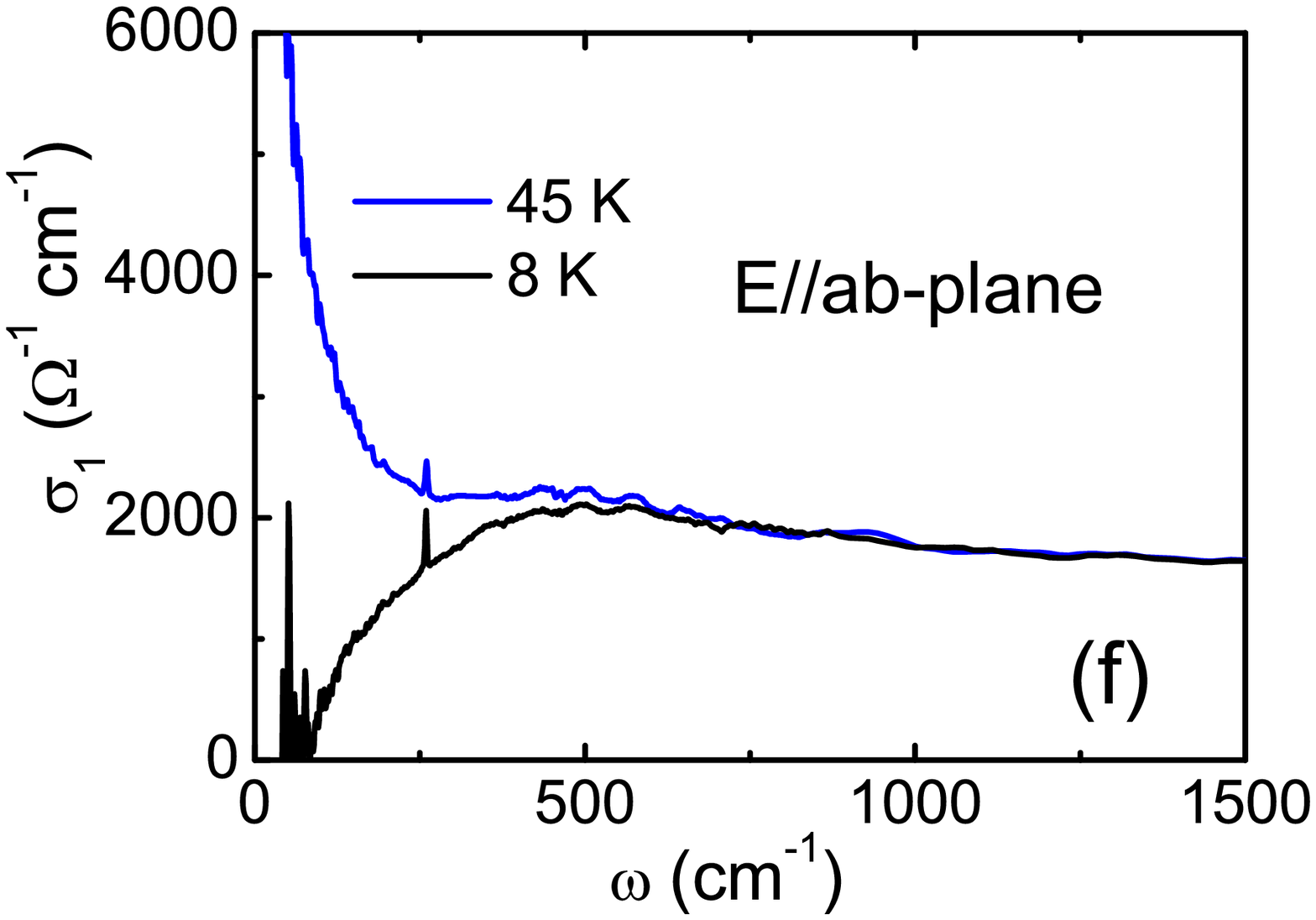}
\caption{(a) optical reflectance R($\omega$) spectra at different
temperatures below 1500 \cm. (b) the conductivity
$\sigma_1(\omega)$ spectra of the sample at different
temperatures. (c) the reflectance R($\omega$) spectra at 8 K and
45 K below 600 \cm. (d) the low-frequency conductivity spectra at
8 and 45 K. For a comparison, the in-plane R($\omega$) and
$\sigma_1(\omega)$ spectra at 8 K and 45 K are shown in (e) and
(f), respectively. The in-plane reflectance and conductivity
spectra show an s-wave superconducting energy gap. By contrast, a
significant residual Drude response is still observed for
\textbf{E}$\parallel$c-axis.}
\end{figure}

The absence of Josephson plasma edge could be taken as an
indication that the 122 iron-pnictides are quasi 3D systems with
the presence of dispersive band along the c-axis. Presence of 3D
Fermi surfaces (FSs) was also suggested from band structure
calculations\cite{GTWang} and a number of other experimental
probes.\cite{Tanatar,Yuan,Martin,Reid,Liu} In particular, recent
{\it ab-initio} LDA+Gutzwiller calculations, where electron
correlations are taken into account beyond LDA, indicated that the
large-size 3D ellipsoid like FS has a dominant Fe-3d$_{3z^2-r^2}$
orbital characteristic\cite{GTWang}. Then the key issue is to
understand why an s-wave superconducing energy gap is clearly seen
from the in-plane optical measurement, whereas a large residual
quasiparticle population is indicated in the c-axis polarized
optical measurement. It has important implication for unravelling
the superconducting pairing in iron-pnictide systems.

In our previous study on undoped parent compounds, we found that
the spin-density-wave gap structure in the c-axis optical spectrum
is significantly different from that in the ab-plane. The
difference was explained naturally by assuming the existence of
two types of FSs in the system: 2D cylinder like FSs and a
large-size 3D ellipsoid like FS.\cite{ZGChen} Here the data could
be again well understood from the same electronic structures, as
schematically shown in Fig. 4.\cite{GTWang} Because the Fermi
velocities of the 2D cylinder-like FSs are essentially within the
ab-plane, the electrons on those FSs would contribute dominantly
to the ab-plane conductivity, while the c-axis conductivity is
mainly contributed by the 3D ellipsoid-like FS. On this basis, the
superconducting energy gaps opened on the 2D cylinder-like FSs
below T$_c$ should be clearly seen by the ab-plane polarized
measurement, but not by the c-axis polarization. Then the c-axis
data could be easily understood if we assume that the
superconducting gap formed on the 3D FS is strongly k-space
dependent, in particular, if node exists in the 3D FS in the
region where the electrons contribute dominantly to the c-axis
optical conductivity.

\begin{figure}[t]
\begin{center}
\includegraphics[width=5cm,clip]{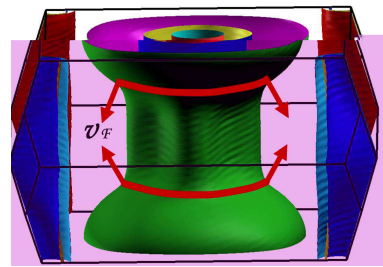}
%\includegraphics[width=3.2in]{fig4.eps}%
%\vspace*{-0.20cm}%
\caption{\label{fig:EuRandSandRho} The Fermi surfaces of
(Ba,K)Fe$_2$As$_2$ from LDA+Gutzwiller calculations\cite{GTWang}.
Besides the 2D cylinder-like FSs, there is an additional 3D FS.
Possible horizontal nodes may exist near the highly flared region
of 3D FS which has maximum contribution to the c-axis
conductivity.}
\end{center}
\end{figure}

It is noted that recent functional renormalization group
calculations on LaFePO superconductors\cite{FWang} indicate that
the 3D hole-like FS centered around ($\pi,\pi,\pi$) with a
dominant Fe-3d$_{3z^2-r^2}$ orbital characteristic is gapless,
although in reality this FS may be absent for LaFePO. Since the 3D
FS in K-doped 122 has the same Fe-3d$_{3z^2-r^2}$
characteristic,\cite{GTWang} furthermore, the FS is not nested
with any other FSs, it is possible that this FS has weak
superconducting strength being induced by the proximity effect or
by the weak mixing of different orbitals.

Our experimental results are consistent with recent c-axis
penetration depth and heat transport
measurements\cite{Martin,Reid,Hirschfeld}, which also suggested
that the nodes contribute specifically to the c-axis transport.
Recent neutron scattering experiment on the same batch of
Ba$_{0.67}$K$_{0.33}$Fe$_2$As$_2$ crystal revealed that the
low-energy spin excitations below T$_c$ displays a sinusoidal
modulation, indicating clear gap at momentum transfer
\textbf{q}$_z$=0 but gapless behavior at
\textbf{q}$_z$=$\pi$.\cite{CLZhang} Presence of nodes in different
regions of FSs was also suggested in theoretical studies,
including nodes on 2D electron type FSs near the zone
corner\cite{Kuroki}, vertical and small segment nodal lines on the
3D hole FS and horizontal nodes on the 3D FS\cite{Graser}. As we
elaborated above, the c-axis optical measurement mainly probes the
3D FS, while the nodes on the 2D electron FSs would have strong
effect on the in-plane conductivity. Since the in-plane optical
data did not reveal substantial residual conductivity in the
superconducting state\cite{Li}, this possibility should be rule
out. Our data suggest that the horizontal node is most likely to
be present, since the vertical node on the 3D FS should also
contribute visibly to the ab-plane conductivity. If the horizontal
nodes exist in the region that contribute dominantly to the c-axis
optical conductivity, for example, in the highly flared region of
3D FS as shown in Fig. 4, meanwhile the gap amplitude changes in
the 3D FS, i.e. being small near the node and relatively large in
the region of k$_z\sim\pm\pi$ where the Fermi velocity has
vanishing c-axis component, then the subtle change in the optical
reflectance spectrum across the T$_c$ is expected.

To conclude, our c-axis polarized optical measurement on the
Ba$_{0.67}$K$_{0.33}$Fe$_2$As$_2$ single crystals revealed that
(1) in contrast to the high-T$_c$ cuprates, no Josephson plasma
edge in R($\omega$) develops below T$_c$; (2) different from the
ab-plane optical response where an s-wave superconducting gap is
clearly observed, the c-axis data only exhibit a small difference
across T$_c$ with the indication of a large residual quasiparticle
population. Our study indicates that the 122 iron-pnictides are
quasi 3D systems with the presence of dispersive band along the
c-axis. Furthermore, there may exist node in the superconducting
gap in the 3D FS in the region which dominates the c-axis optical
conductivity. We suggest that a horizontal node at the highly
flared region of 3D FS is more consistent with our experimental
observation.

\begin{acknowledgments}
We thank X. Dai, H. Ding, Z. Fang, J. P. Hu, D. H. Lee, X. G. Wen
and L. Yu for useful discussions. This work is supported by the
National Science Foundation of China, the Chinese Academy of
Sciences, the 973 project of the Ministry of Science and
Technology of China, and the US Department of Energy, Division of
Materials Science, Basic Energy Sciences, through DOE
DE-FG02-05ER46202.

\end{acknowledgments}

\end{document}